# O/IR Polarimetry for the 2010 Decade (SSE): Science at the Edge, Sharp Tools for All

A Science White Paper for the
**Stars and Stellar Evolution (SSE) Science Frontiers Panel** of the
Astro2010 Decadal Survey Committee

Lead Authors:

Jennifer L. Hoffman
Physics and Astronomy
University of Denver
2112 E. Wesley Ave.
Denver, CO 80208
(303) 871–2268
jennifer.hoffman@du.edu

Dean C. Hines
Space Science Institute
New Mexico Office
405 Alamos Rd.
Corrales, NM 87048
(505) 239-6762
hines@spacescience.edu

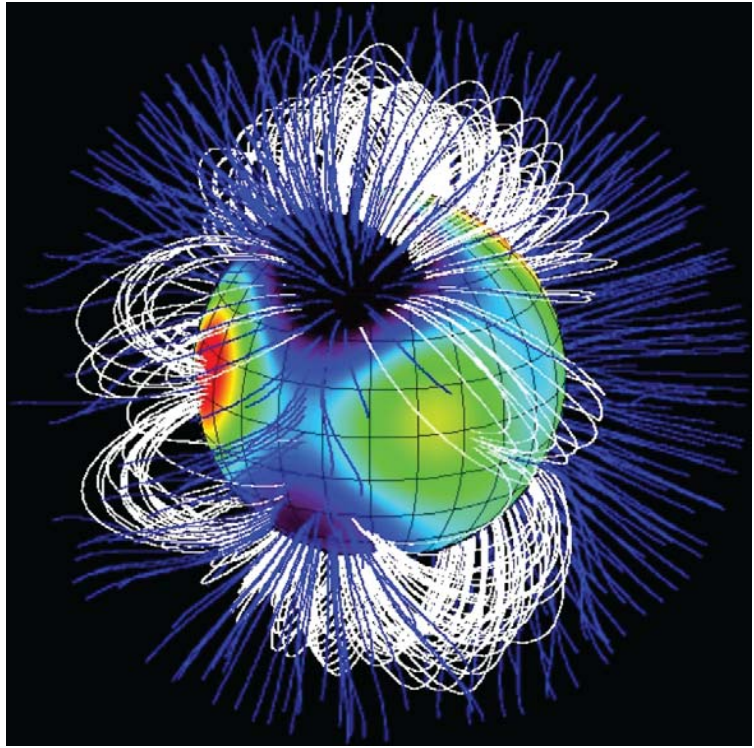

*Surface magnetic field of the B0.5V star τ Sco, which shows a strongly non-dipolar field topology (Donati et al. 2006).*

*Contributors and Signatories*

| | |
|---|---|
| Andy Adamson | UKIRT, JAC, Hilo |
| B-G Andersson | Stratospheric Observatory for Infrared Astronomy |
| Karen Bjorkman | University of Toledo |
| Ryan Chornock | University of California, Berkeley |
| Dan Clemens | Boston University |
| James De Buizer | SOFIA, NASA Ames |
| Nicholas M. Elias II | Universität Heidelberg; MPIA |
| Richard Ignace | East Tennessee State University |
| Terry Jay Jones | University of Minnesota |
| Alexander Lazarian | University of Wisconsin |
| Douglas C. Leonard | San Diego State University |
| Antonio Mario Magalhaes | University of Sao Paulo, Brazil |
| Marshall Perrin | UCLA |
| Claudia Vilega Rodrigues | Inst. Nac. De Pesquisas Espaciais, Brazil |
| Hiroko Shinnaga | CalTech |
| William Sparks | STScI |
| Lifan Wang | Texas A&M University |

*Overview and Context: Polarimetry as a cross-cutting enterprise*

Photometry, spectroscopy, and polarimetry together comprise the basic toolbox astronomers use to discover the nature of the universe. Polarimetry established the Unified Model of AGN and continues to yield unique and powerful insight into complex phenomena. Polarimetry reveals the elusive magnetic field in the Milky Way and external galaxies, allows mapping of features of unresolved stars and supernovae, uncovers nearby exoplanets and faint circumstellar disks, and probes acoustic oscillations of the early universe.

Polarimetry is practiced across the full range of accessible wavelengths, from long wavelength radio through gamma rays, to provide windows into phenomena not open to photometry, spectroscopy, or their time-resolved variants. At some wavelengths, the U.S. leads the world in polarimetric capabilities and investigations, including ground-based radio, through the VLA and VLBA. At other wavelengths, the U.S. is currently competitive: in sub-mm the CSO and the JCMT have historically pursued similar science problems. In ground-based O/IR, the situation is considerably worse, with no optical or NIR polarimeters available on Gemini (Michelle is MIR only) or any NOAO-accessed 4 m telescope, as the table below shows. Over the past decade and more, Canadian and European astronomers have enjoyed unique access to state-of-the-art polarimeters and have used this access to vault far past the U.S. in many science areas.

| Telescope | Aperture | Instrument | Waveband | Polar. Mode | U.S. Access ? |
|---|---|---|---|---|---|
| IRSF (SAAO) | 1.4m | SIRPOL | NIR | Imaging | No |
| Perkins (Lowell) | 1.8m | Mimir, PRISM | NIR, Optical | Imaging | Private |
| HST | 2.4m | WFPC2, ACS, NICMOS | Optical, Optical, NIR | Imaging | Yes |
| Nordic Optical | 2.5m | TURPOL | Optical | Photopol | No |
| MMT | 6.5m | MMTPOL | NIR | Imaging | Private |
| LBT | 2x8.4m | PEPSI | Optical | Spectropol | Private |
| Gemini | 8m | Michelle | MIR | Imaging | Yes |
| Keck | 10m | LRIS | Optical | Spectropol | Private |
| GTC | 10m | CanariCam | MIR | Imaging | No |

In space, NICMOS, ACS, and WFPC-2[1] on HST have permitted imaging polarimetry down to 0.1% precision, and may represent the most general purpose O/IR access for U.S. astronomers. Neither the Spitzer Space Telescope nor JWST provides, or will provide, any polarimetric capability. The dwindling U.S. access to this crucial third leg of the light analysis tripod has also become self-fulfilling, as students receive little exposure to polarimetric techniques and scientific advances as the number of practitioners able to teach students declines.

Nevertheless, polarimetric studies in O/IR have already revealed a great deal about star and planet formation processes, stars and their evolution, the structure of the Milky Way, and the nature and origin of galaxies and their active nuclei – details which cannot be discovered using pure photometric or spectroscopic methods. For example, NIR imaging polarimetric studies by the SIRPOL group (e.g., Tamura et al. 2007) reveal the details of the magnetic fields lacing nearby, star-forming molecular cloud cores and the embedded reflection nebulae and disks associated with their newly formed stars. Further, the race to find and image exoplanets will use polarimetry internal to the two extreme adaptive optics coronagraphs now under construction: SPHERE/ZIMPOL (for the VLT: Joos 2007) and GPI (for Gemini: Macintosh et al. 2006).

---

[1] NICMOS and much of ACS are currently off-line until Servicing Mission 4 (SM4); WFPC-3 will replace WFPC-2 but will have no polarimetric capability.

Theoretical efforts have recently advanced our understanding of the origin of dust grain alignment by demolishing old theory and offering new, testable predictions. The new paradigm of radiative aligned torques (e.g., Lazarian & Hoang 2007; Hoang & Lazarian 2008) is removing old doubts concerning magnetic alignment while leaving a wide parameter space open for observational testing and future theoretical refinement.

Polarimetric modeling has also entered a modern age, one characterized by huge model grids, spectacular dynamic ranges, and closer coupling to observational measurements. The active debate between the relative importance of magnetohydrodynamics (e.g., Li et al. 2004) vs. pure hydrodynamics (e.g., Padoan & Nordlund 2002) is shining new light into the nature of magnetic fields in the ISM and in star formation. Meanwhile, radiative transfer modeling guides detailed interpretation of complex polarized line profiles from aspherical stellar winds (e.g., Harries 2000) and supernovae (e.g., Höflich 2005; Kasen et al. 2006; Hoffman 2007). This upcoming decade is sure to see careful, detailed comparisons between high-resolution model simulations and observational data that will lead to sharp new insights into many astrophysical scenarios.

The promise evident in the new, niche polarimetric instruments and the surveys they will perform will drive cutting-edge science in the upcoming decade. Yet teasing out answers to many key questions requires open community access to general-purpose, precision polarimeters on large telescopes, as well as opportunities for student training.

### *Example Polarimetry Science Areas for the next decade*

Polarimetric studies are crucial for many areas of stellar astrophysics. Here we highlight several that show particular promise for transforming the field in the coming decade.

- <u>Imaging the surfaces of stars in polarized light to unveil atmospheric phenomena</u>

Stars represent the bedrock of astronomy. Understanding their astrophysical processes from core to photosphere to circumstellar environment is critical for research in fields as diverse as star/planet formation, galactic dynamics, galaxy formation, and cosmology. Not long ago, the Sun was the only star whose surface we could study, but recent interferometric advances now allow us to create direct, detailed images of the surfaces of other stars. For example, Monnier et al. (2007) resolved the surface of Altair and found that the temperature at its pole is higher than that at the equator. This suggests that the spectral types of stars may depend on our viewing angle, a profound paradigm shift that has wide-ranging implications for all areas of astronomy. Further exploration of "extrasolar" stellar surfaces will be a crucial astrophysical advancement of the next decade.

The emerging technique of optical interferometric polarimetry (OIP) enables stellar surface imaging in polarized light (Elias et al. 2008). OIP breaks the symmetry generally observed in "normal" main-sequence stars, yielding detailed views of the constituents and scattering processes within their atmospheres. Using OIP, we will be able to image the unusual surface features of Ap/Bp stars and thereby probe in unprecedented detail the kilogauss magnetic fields and peculiar chemical abundances that characterize these objects (Shulyak et al. 2008). We will be able to construct detailed maps of the magnetic field structures in M dwarfs, thus illuminating the action of the dynamos that power the surface activity in these and other convective stars (Donati et al. 2008). We will surely discover unexpected phenomena that challenge our current picture of the processes that drive stellar evolution and shape the circumstellar environments of stars of all types. The synergy between polarimetry and interferometry provided by OIP observations promises to open a new era of stellar astrophysics, in which we observe other stars with the same level of detail as we do our Sun.

- <u>Probing the structure and dynamics of circumstellar disks in a range of environments</u>

Far from being isolated objects, stars of many types and all evolutionary stages interact with their surroundings via circumstellar disks. These disks affect both the formation and the evolution of their host stars throughout their lives, and sculpt the complex structures seen and inferred in star-forming environments, bipolar planetary nebulae, luminous blue variable outbursts, supernovae, and gamma-ray bursts (GRBs). Understanding the characteristics of these disks is crucial to unraveling questions about their origins and pinpointing their effects on stellar evolution.

When a source is unresolved, polarimetric observations provide particularly powerful disk diagnostics and constraints not easily obtainable by any other means. Using spectropolarimetry, we can directly measure a faraway disk's spatial orientation and constrain its geometry (e.g. Quirrenbach et al. 1997; Wood et al. 1997). In addition, because polarized light samples the disk's opacity, spectropolarimetric measurements directly determine physical conditions such as temperature, ionization state, density, and velocity structure inside a circumstellar disk, whether or not the disk is resolved. In the near future, spectropolarimeters on large telescopes will open new arenas of inquiry into the nature of circumstellar disks in a variety of new environments, such as the low-metallicity regions of the LMC and SMC (Wisnewski & Bjorkman 2006)

Most importantly for the next decade, the combination of spectropolarimetry with the established tools of photometry, spectroscopy, and interferometry at many wavelengths will bring about the most detailed understanding of circumstellar disk structures yet possible. New queue-scheduled spectropolarimeters will provide access to the time domain and broaden our understanding of disks' variability on a range of time scales. Emerging 3-D non-LTE radiative transfer simulations will aid interpretation of these complex data and allow us for the first time to develop detailed, well-constrained, realistic physical models of circumstellar disks. In an example of the powerful synergy that will characterize the new era of disk studies, Carciofi et al (2009) combined quantitative radiative transfer modeling with long time-baseline photometric, spectroscopic, interferometric, and spectropolarimetric data to confirm the presence of a one-armed spiral density wave within the circumstellar disk of the classical Be star $\zeta$ Tauri. This work presages the sophisticated results that will become commonplace in the next ten years with the addition of spectropolarimetry and its associated analysis tools to our diagnostic toolbox.

- <u>Deciphering the nature of stellar winds and magnetic fields</u>

Many problems in the arena of extended stellar atmospheres and winds can only be addressed through polarimetric studies. Polarimetry is an essential tool for probing clumpy wind structures in luminous blue variables (Davies et al. 2005), illuminating aspherical mass loss in single stars and binary systems (e.g., Hoffman et al. 1998; Villar-Sbaffi et al. 2005; Ignace et al. 2009), exploring the origins of the asphericities in post-AGB stellar winds and planetary nebulae (e.g., Trammell et al. 1994, 1995; Ueta et al. 2007), and characterizing the accretion processes in polars and magnetic cataclysmic variables (e.g., Butters et al. 2009).

In recent years, spectropolarimetry has transformed our understanding of massive stellar atmospheres and winds by allowing detailed mapping of stellar magnetic fields using the Zeeman effect (e.g., Ignace & Gayley 2003; Alecian et al. 2008a,b; Hubrig et al. 2008). Zeeman analysis of the young, massive star τ Sco revealed a strongly non-dipolar surface magnetic field (Donati et al. 2006; see cover-page figure). In another example, using Chandra X-ray data paired with MHD simulations of magnetically channeled wind shocks on $\theta^1$ Ori C, Gagné et al. (2005) constructed a comprehensive model of the properties of the X-ray plasma in this nebula-shaping O star. The next decade will see widespread proliferation of such detailed studies. In addition, new polarimetric tools such as the Hanle effect (Ignace et al. 2004), atomic and dust-grain alignment signatures (Lazarian & Hoang 2007; Nordsieck 2008), and X-ray polarimetry (McNamara et al. 2008) will transcend current observational limitations and forge a new picture of magnetic stellar winds. In particular, quantifying the magnetic field structures of stars in late evolutionary stages will be crucial to answering current questions about the links between supernovae/GRBs and their massive progenitors (e.g., Davies et al. 2005; Vink 2007).

*Exceptional Discovery Potential Area:*
*The Sources and Implications of Supernova Asymmetries*

Over the past decade, the emerging field of supernova spectropolarimetry has revealed increasing complexities in the cosmic explosions that were once assumed to be homogeneous and spherical (see Wang & Wheeler 2008 for a review and comprehensive list of references). Supernovae of all types show large line polarization signatures that point to significant asphericities in their ejecta. Spectropolarimetric measurements of supernovae have been relatively rare in the past, but over the next ten years, improving statistics and higher-quality measurements promise to yield unprecedented insights into the nature of these and related transient energetic phenomena.

In core-collapse supernovae, observed changes in polarization across key spectral lines show that the explosion mechanism is highly aspherical in nature; one explanation posits jet-like flows that precess over time and may be related to GRB jets (e.g., Trammell et al. 1993; Cropper et al. 1988; Leonard et al. 2006; Maund et al. 2007a). Spectropolarimetry can also constrain the nature of core-collapse progenitors by characterizing the distribution of elements within the ejecta and in the surrounding, previously ejected CSM (Wang et al. 2001; Hoffman et al. 2008). In addition, spectropolarimetry probes the links between supernovae and the high-energy transient phenomena of GRBs and X-ray flashes (Kawabata et al. 2003; Maund et al. 2007b). Future detailed spectropolarimetric studies of core-collapse supernovae will clarify the relations between object types and illuminate their three-dimensional nature and behavior.

In thermonuclear supernovae, strong line polarization prior to maximum light reveals the existence of large-scale asymmetries in the outer ejecta that cannot be explained by current explosion models (e.g., Kasen et al. 2003; Chornock & Filippenko 2006; Wang et al. 2006). Spectropolarimetry also yields clues to the nature of peculiar and "hybrid" Type Ia-like objects that challenge our current classification system (e.g., Wang et al. 2004; Chornock et al. 2006). Spectropolarimetric observations of thermonuclear supernovae are difficult to obtain, but these data contain important information about the nature of Type Ia explosions that may have significant cosmological implications due to their use as standard candles.

Over the next decade, the greatest advances will come from a combination of high-resolution spectropolarimetry with advanced 3-D hydrodynamical and radiative transfer tools that can simulate the strongly aspherical geometries characterizing our current understanding of supernovae. Observations across a wide variety of wavelength regimes and with an emphasis on the time-dependent behavior of these transient objects promise to reveal telling details of the aspherical nature of the explosions and allow us to probe the physical characteristics of both their immediate circumstellar neighborhoods and their larger galactic environments. Meanwhile, the increasing sophistication of numerical simulations will enable robust interpretation of spectropolarimetric signatures and the reliable linking of observations with theoretical models.

## *What is Needed to Meet the Science Goals within the Decade*

What key observations, theory, and instrumentation are needed to achieve the science goals within the next decade? Our evaluation of the upcoming opportunities, challenges, and technical readiness leads to the following recommendations:

1. <u>Build precision polarimetric capability into new O/IR instruments for large telescopes and space missions. Design polarimetry in from the beginning, not as "add-ons"</u>.
   - New "niche" instruments, such as GPI, SPHERE/ZIMPOL, and HiCIAO on Subaru rightly exploit polarimetry to meet their exoplanet objectives, but general-purpose instruments with polarimetric capability are lacking at virtually all large, open-access US telescopes, whether ground-based, airborne, or space-based.
   - Key science questions cannot be answered unless US astronomers have access to precision (photon-noise limited) polarimetric capability. To retain precision capability, polarimetric capabilities must be a considered in the initial design of the instruments, not as a later "add-on".

2. <u>Encourage polarimetric surveys with LSST and synoptic and survey capabilities on intermediate-size telescopes</u>.
   - Since much of the sky has never been explored polarimetrically, a polarimetric LSST survey promises new frontiers in stellar astrophysics[2].
   - Precision interstellar polarization measurements are key to reliable observation of intrinsic spectropolarimetric signatures of stars and supernovae.

3. <u>Develop polarimetric O/IR synoptic and survey capabilities on intermediate-size telescopes to probe the time evolution of stars and supernovae, as well as to train students in instrumentation and polarimetric observations</u>.
   - To be able to compete scientifically, we must invest in the next generation of young astronomers who will use polarimetry as a powerful tool in their light analysis toolbox and who will understand polarimetric light analysis well enough to guide future instrument development.
   - Synoptic polarimetric observational data sets are crucial to understanding stellar phenomena with complex geometry and/or time evolution.

---

[2] See the All-Sky Polarimetry Survey White Paper submitted to the LSST Consortium (http://astroweb.iag.usp.br/~mario/).

- Obtaining ground-based calibrating polarimetric observations is vital to the calibration of existing and future space-based polarimeters.

4. <u>Develop sophisticated computational and theoretical models to help interpret increasingly complex spectropolarimetric data.</u>
    - Spectropolarimetric observations are notoriously complex, but sophisticated computational tools allow us to take advantage of the rich scientific content of these data.
    - The field of stellar polarimetry is currently driven by observation. Rapid advances in numerical modeling techniques are necessary to enable a full understanding of the data and make predictions that can be tested with future observations.

*Final Thought*

The U.S. astronomical community has lost opportunities to advance key science areas as a result of down-selects of instrument capabilities or lack of will to commission polarimetric modes on instruments. The investment is minor, the expertise is available in the community, and the rewards are tangible. We are excited by the recent momentum favoring polarimetric studies and capabilities and believe the upcoming decade will see the various polarimetric techniques together become a strong, necessary component of astronomers' light analysis toolbox.

*Bibliography and References*